\documentclass[aps,twocolumn,amsmath,showkeys,prl,superscriptaddress
]{revtex4-1}

\usepackage{dcolumn}
\usepackage{bm}
\usepackage[T1]{fontenc}
\usepackage{amsmath}
\usepackage{graphicx} 
\usepackage{tabularx}
\usepackage{multirow}
\usepackage{rotating} 
\usepackage{makecell}
\usepackage{amssymb}
\usepackage[outercaption,wide]{sidecap}
\usepackage{hyperref}
\hypersetup{colorlinks=true, linkcolor=blue, citecolor=blue, urlcolor=blue}

\usepackage[normalem]{ulem}
\usepackage[usenames, dvipsnames]{xcolor}
\newcommand\blue[1]{{\color{black}#1}}

\begin{document}
\title{Iron-based superconductivity extended to the novel \blue{silicide} LaFeSiH
}
\author{F. Bernardini}
\affiliation{CNR-IOM-Cagliari and Dipartimento di Fisica, Università di Cagliari, IT-09042 Monserrato, Italy}
\author{G. Garbarino}
\affiliation{European Synchrotron Radiation Facility, 6 rue Jules Horowitz, BP 220, 38043 Grenoble, France}
\author{A. Sulpice}
\author{M. N\'{u}ñez-Regueiro}
\affiliation{CNRS, Universit\'e Grenoble Alpes, Institut N\'eel, 38042 Grenoble, France}
\author{E. Gaudin}
\author{B. Chevalier}
\affiliation{CNRS, Univ. Bordeaux, ICMCB, UPR 9048, F-33600 Pessac, France}
\author{M.-A. M{\'e}asson}
\affiliation{CNRS, Universit\'e Grenoble Alpes, Institut N\'eel, 38042 Grenoble, France}
\author{A. Cano}
\email{andres.cano@cnrs.fr}
\author{S. Tencé}
\affiliation{CNRS, Univ. Bordeaux, ICMCB, UPR 9048, F-33600 Pessac, France}
\date{\today}

\begin{abstract}
We report the synthesis and characterization of the novel \blue{silicide} LaFeSiH displaying superconductivity \blue{with onset at 11~K}. We find that this pnictogen-free compound is isostructural to LaFeAsO, with a similar low-temperature tetragonal to orthorhombic distortion. Using density functional theory we show that this system is also a multiband metal in which the orthorhombic distortion is likely related to single-stripe antiferromagnetic order. Electrical resistivity and magnetic susceptibility measurements reveal that these features occur side-by-side with superconductivity, which is suppressed by external pressure.
\end{abstract}

\maketitle

Iron-based superconductors (Fe-based SCs) provide an unprecedented playground for the investigation of high-$T_c$ superconductivity. These systems belong to a huge family of compounds and recurrently display the following key features (see e.g. \cite{martinelli-crp,*bascones-crp,*inosov-crp,*bohmer-crp,*roekeghem-crp,*rullier-crp,*zapf-crp,*hirschfeld-crp,gallais-crp} for recent reviews). From the structural point of view, they have Fe$X$ layers in which the Fe atoms form a square lattice that is sandwiched between two $(\sqrt{2}\times\sqrt{2})R45^\circ$ shifted lattices
of $X$ ($=$ P, As, Se, Te, S).
This leads to a quasi-2D multiband Fermi surface that mainly originates from the Fe 3$d$ orbitals. In addition, the parent compounds often display anti-ferromagnetic (AFM) order 
\blue{inducing a lattice distortion \cite{yildrim-prl08,cano-prb10,cano11} that is generally pre-empted by the so-called nematic transition \cite{gallais-crp}.}
In the prototypical case of $R$FeAsO ($R=$ rare earth), for example, this specifically corresponds to single-stripe AFM order [also called $(\pi,0)$ order] and a square-to-rectangular distortion of the Fe layers. These features advocate for the so-called $s_{\pm}$ superconducting gap symmetry
and a spin-fluctuation pairing. This superconductivity can be induced by carrier doping resulting from either chemical substitutions or physical pressure \cite{gg-prb-2008,*gg-prb-2011}. From a more methodological point of view, the electronic band structure and the magnetic orders found in Fe-based SC can be reasonably well described by means of DFT-based calculations as the electronic correlations often remain relatively weak. 

The search for novel Fe-based SC has been naturally extended to systems in which the Fe$X$ layer contains group IV elements, \blue{in particular the non-toxic Ge}. Thus, MgFeGe for example has been identified as isostructural and isoelectronic to the LiFeAs compound but displaying no superconductivity \cite{welter-98,hosono-MgFeGe-prb-12,roser-MgFeGe-prb13}. Evidence for superconductivity below 2~K has recently been reported in YFe$_2$Ge$_2$ \cite{grosche-pss14,grosche-prl16}, although whether bulk superconductivity is actually realized in this material is currently under debate \cite{canfield-15}. On the other hand, hydrogen substitution has proven to be a particularly interesting route to induce superconductivity in these systems. Specifically, the carrier doping limit of the original F substitution in LaFeAsO has been surpassed with hydrogen, thus revealing a two-dome superconductivity \cite{hosono-natcomm12} together with a second magnetic phase in the additional parent compound LaFeAsO$_{0.5}$H$_{0.5}$ \cite{hosono-nphys14}. In this paper, we report the synthesis, crystal structure and physical properties of the novel compound LaFeSiH. In addition, we study the behavior of the system under hydrostatic pressure and perform density-functional-theory (DFT) calculations to obtain the corresponding electronic band structure and magnetic ground state of this compound. Thus, we find that the new compound LaFeSiH  is isostructural, isoelectronic and also ``isomagnetic'' to the 1111 family of Fe-based SC. Most importantly, we find superconductivity \blue{with onset at 11~K} in this pnictide- and chalcogenide-free system. 

\begin{figure}[tbp!]
\centering
\includegraphics[width=0.275\textwidth]{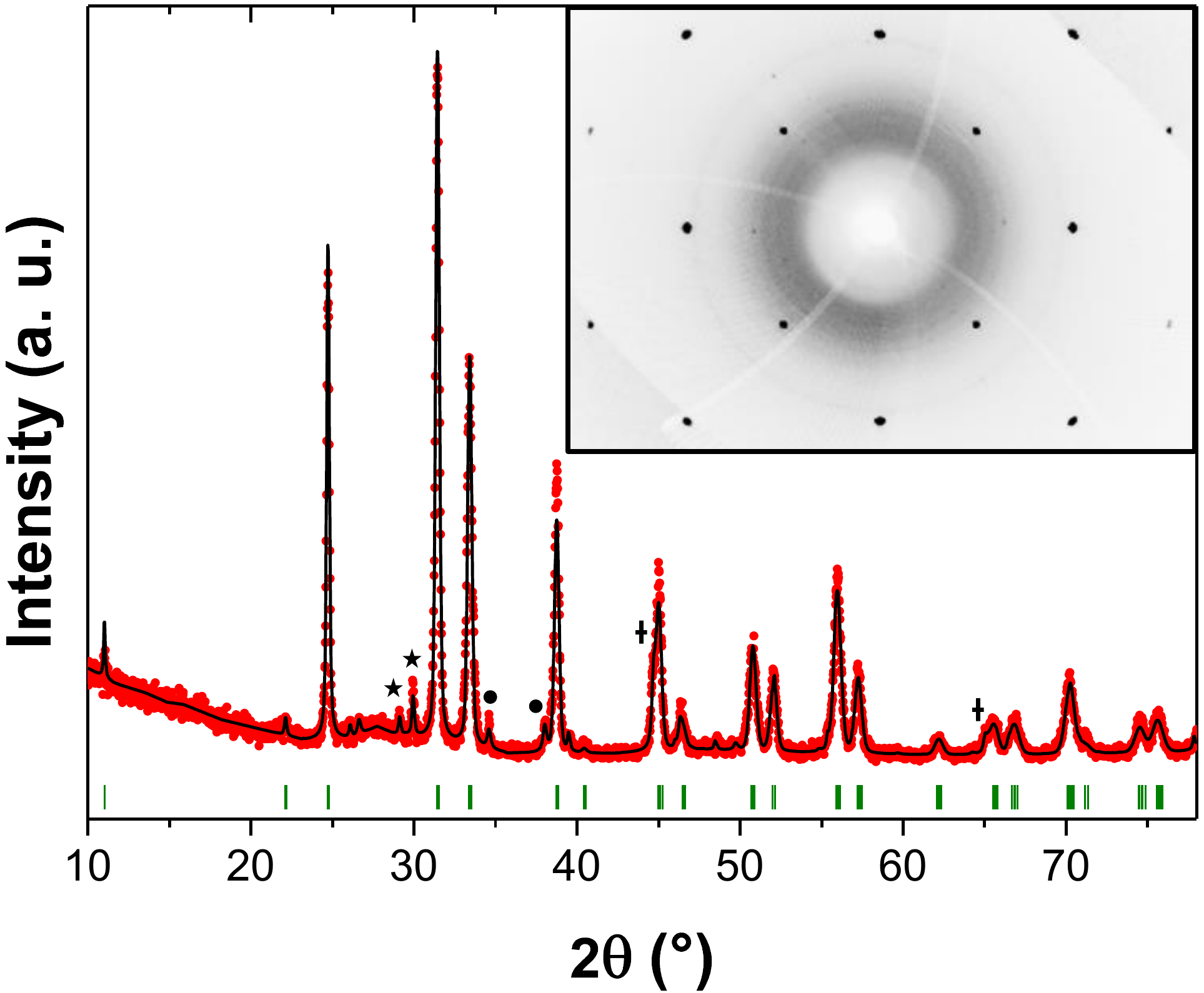}
\includegraphics[width=0.2\textwidth]{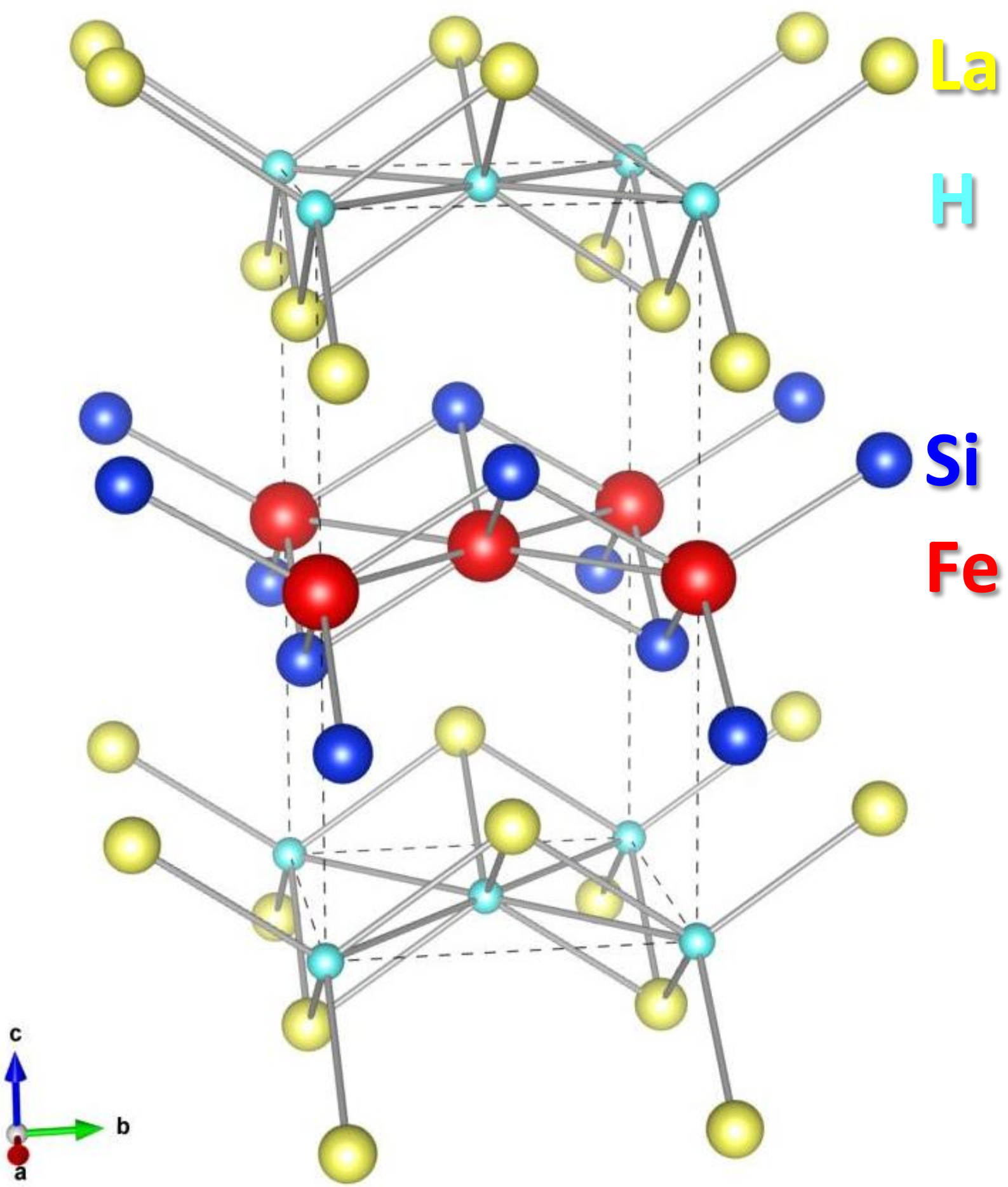}\caption{(Left) Room temperature x-ray powder diffraction pattern of LaFeSiH (p1). Ticks indicate the Bragg peaks of the $P4/nmm$ structure of the main LaFeSiH phase while stars, crosses and circles indicate extra peaks due to the secondary phases La$_2$O$_3$, $\alpha$-Fe and La(Fe,Si)$_{13}$H$_x$ respectively that could not be completely eliminated by annealing. 
The $(hk0)$-plane of the LaFeSiH single-crystal diffraction pattern at room temperature is shown in the inset \blue{(s1)}. (Right) Ball-and-stick model of the crystal structure of LaFeSiH. 
\label{ball-and-stick}}
\end{figure}

\paragraph{Crystal structure.---} We synthesized the \blue{silicide hydride} LaFeSiH as described in the Supplemental Material, \blue{and singled out two micrometric single crystals (s1 and s2) from two powder samples (p1 and p2).}
The compound is stable in air, and its room-temperature crystal structure corresponds to the tetragonal space group $P4/nmm$ with the structural parameters shown in Table \ref{structure}. 
This structure was determined from the Rietveld refinement of the x-ray \blue{and neutron} powder diffraction pattern and more accurately from single-crystal x-ray diffraction measurements (see Fig.~\ref{ball-and-stick} and Supplemental Material). 
Compared to LaFeSi \cite{welter-jac98}, the new compound displays a variation of the lattice parameters that is highly anisotropic: $a = 4.098 \to 4.027$ {\AA} and $c = 7.133 \to 8.014 $ \AA. This variation is due to the hydrogen insertion at the $2b$ Wyckoff position, as observed in other CeFeSi-type hydrogenated intermetallics \cite{tence-jap09,*chevalier09,*tence10}. This insertion is confirmed from \blue{both} the single-crystal x-ray diffraction pattern \blue{and the neutron diffraction data that, in addition, reveals the full occupancy of the H atoms at the HLa$_{4/4}$ tetrahedra (see Fig. \ref{ball-and-stick} and Supplemental Material) thus confirming the LaFeSiH stoichiometry}. Additionally, we performed DFT calculations \blue{that reveal that this}
is a very stable position for the H atom. Specifically, the overall $P4/nmm$ structure is found to have a well-defined phonon spectrum with H-related optical phonons of frequencies greater than $100.5$ meV (see also \cite{yildrim17}). Thus, we conclude that the room-temperature crystal structure of LaFeSiH is essentially the same structure of the 1111 Fe-based SCs (e.g. LaFeAsO) as illustrated in Fig. \ref{ball-and-stick}.     

We also determined the low-temperature crystal structure by performing additional x-ray synchrotron experiments at 15 K. The refinement of the corresponding structure revealed a low-temperature phase of LaFeSiH with space group symmetry \blue{$Cmme$}. The structural parameters of this phase are in Table \ref{structure}. The orthorhombic distortion is rather small, which is again similar to the observed in Fe-based SCs.

\begin{table}[t!]
\footnotesize 
\begin{tabular}{ccccc}
\multicolumn{5}{l}{
293K - $P4/nmm$ (\#129, origin 2)}\\ 
\multicolumn{5}{l}{
$a = 4.0270(1)${\AA} $c=8.0374(8)$\AA}\\
\hline \hline 
 & Wyckoff pos.  & $x$ & $y$ & $z$ \\
\hline 
La & $2c$ & 1/4 & 1/4 & 0.6722(1) \\
Fe & $2a$ & 3/4 & 1/4 & 0 \\
Si & $2c$ & 1/4 & 1/4 & 0.1500(5) \\
H & $2b$ & 3/4 & 1/4 & 1/2 \\
\hline \hline
\\
\multicolumn{5}{l}{
15K - \blue{$Cmme$} (\#67)}\\ 
\multicolumn{5}{l}{
$a = 5.6831(6)${\AA} $b = 5.7039(6)${\AA}   $c=7.9728(6)$\AA}\\
\hline 
 & Wyckoff pos.  & $x$ & $y$ & $z$ \\
\hline
La & $4g$ & 0 & 1/4 & 0.1747(3) \\
Fe & $4b$ & 1/4 & 0 & 0 \\
Si & $4g$ & 0 & 1/4 & 0.655(1) \\
H & $4a$ & 1/4 & 0 & 0 \\
\hline \hline 
\end{tabular} \caption{Structure parameters of LaFeSiH. (293K) The atomic positions were obtained from single-crystal x-ray diffraction \blue{(s1)}
while the cell parameters from the powder. 
\blue{The high quality of the single crystal enables the identification of the $2b$ H-atom position already from its x-ray diffraction pattern (see Fig. \ref{ball-and-stick}). This is further confirmed from the neutron powder diffraction data (see Supplemental Material).} 
(15K) Structure parameters obtained from Rietveld refinements of x-ray synchrotron \blue{powder data (p1)}.  
\label{structure}}
\end{table}

\begin{figure*}
\includegraphics[width=0.99\textwidth]{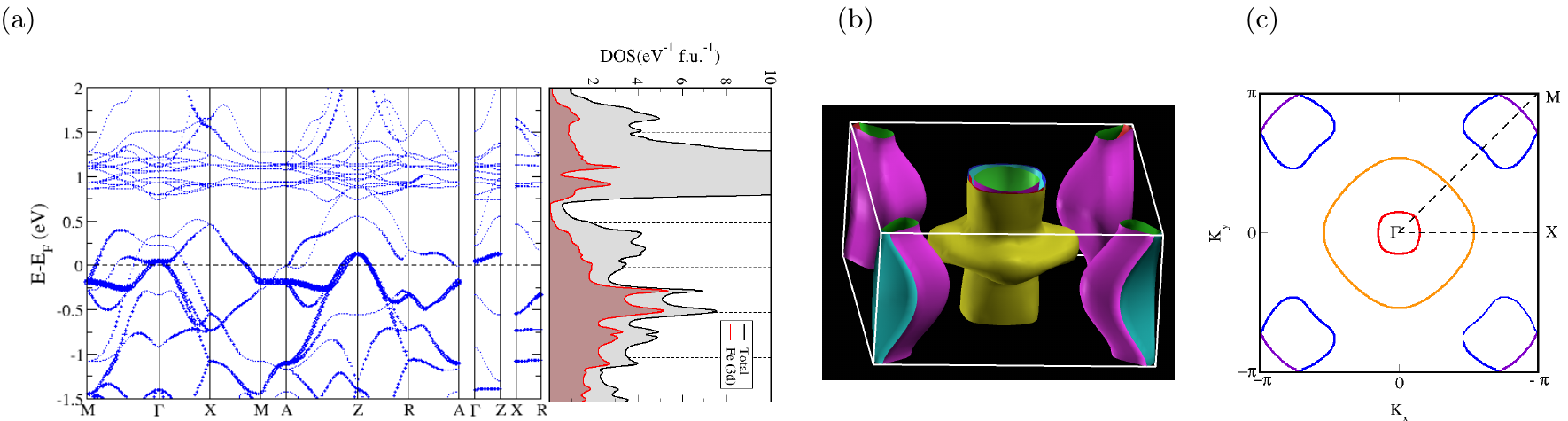}
\caption{(a) Band structure and total DOS and PDOS of the Fe-3$d$ states. (b) Three dimensional view of the Fermi surface and (c) cross section in the $k_z=0$ plane. 
\label{band0}}
\end{figure*}

\paragraph{Electronic and magnetic structure.---} We first computed the electronic structure of LaFeSiH assuming a non-magnetic ground state (see Supplemental Material for details). We took the experimental low-temperature structural parameters as input. However, since the orthorhombic distortion is very small, we averaged the $a$ and $b$ parameters and considered an effective tetragonal structure. 
In Fig. \ref{band0} (a) we show the calculated electronic density of states (DOS) and the partial DOS (PDOS) projected on the Fe $3d$ orbitals. As we can see, the system has a metallic nature with a DOS at the Fermi level that is largely dominated by the Fe-$3d$ orbitals. The corresponding band structure is shown in Fig. \ref{band0} (a), where the size of the symbols reflects the PDOS of the $3(d_{xz} + d_{yz})$ Fe orbitals. 
This structure reveals the presence of several bands crossing the Fermi energy, which renders the system a multiband character. Specifically, we have two hole bands centered at the $\Gamma$ point and two additional bands around the $M$ point. In addition, there is a third band centered at the $\Gamma$ point below the Fermi energy but very close. The resulting Fermi surface is shown in Fig. \ref{band0} (b) and (c). This Fermi surface has a resemblance to the characteristic Fermi surface of the Fe-based SCs. The central lobe originating from the outer hole band and the belly of the otherwise cylindrical electron sheets represent the main differences. 

The nesting properties of the LaFeSiH Fermi surface are limited compared to most of Fe-based SCs. Yet this system  can develop some type of magnetic order. Thus, we performed additional DFT calculations in which different magnetic states were considered. Specifically, we study the tendency of the non-magnetic tetragonal structure towards ferromagnetic (FM), single-stripe AFM, double-stripe AFM and checkerboard AFM ordering of the Fe spins. The energy difference with respect to the non-magnetic state and the corresponding magnetic moments are reported in Table \ref{table.energiesP=0}. We find that the single-stripe AFM order reduces considerably the total energy of the system, and therefore is expected to be the ground state of the system. This explains the orthorhombic distortion we observed experimentally, as this is expected to be a by-product of the single-stripe AFM order triggered by magnetostructural coupling \cite{yildrim-prl08,cano-prb10,cano-prb12}.

\begin{figure}[b!]
\centering
\includegraphics[width=0.35\textwidth]{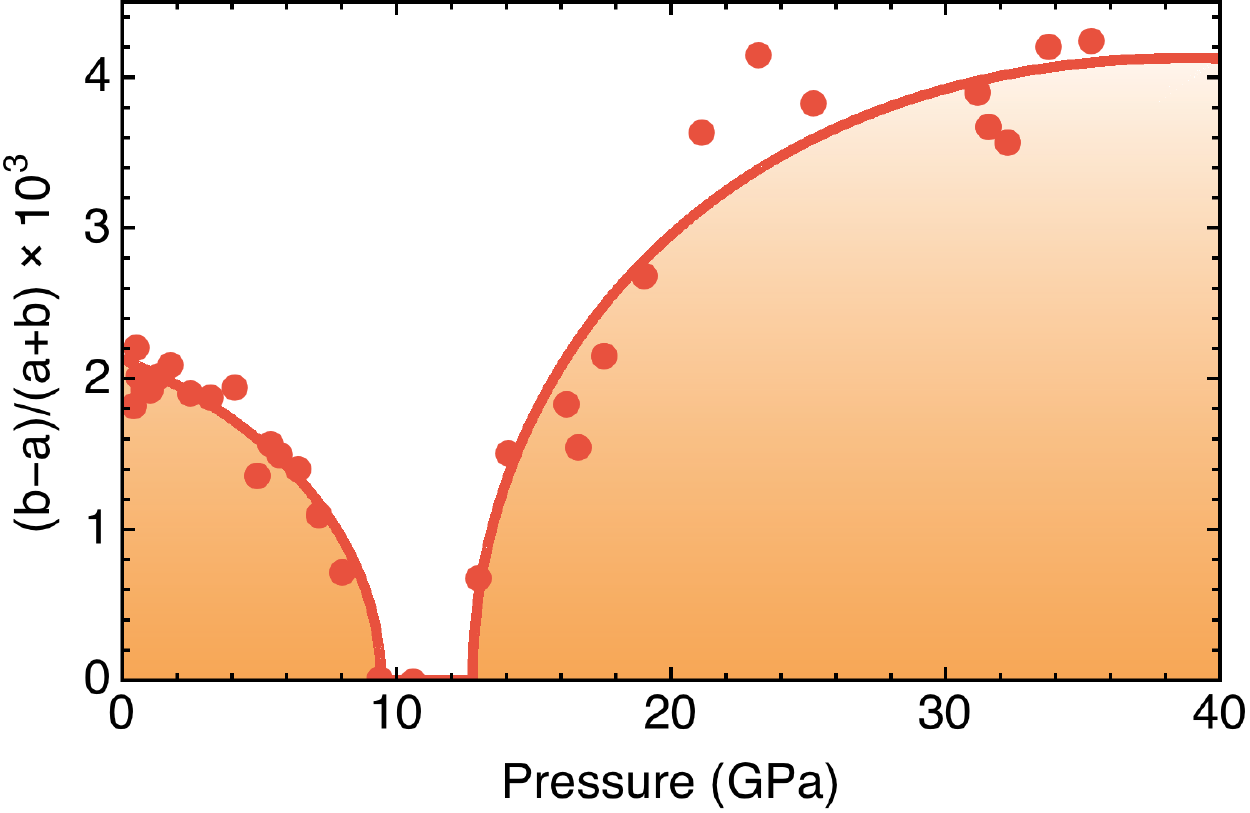}
	\caption{\label{distortionvsP} 
	Relative difference between the $a$ and $b$ lattice parameters of LaFeSiH at 15K as a function of pressure. Lines are guides to the eye.}
\end{figure}

\paragraph{Crystal and magnetic structure under pressure.---} Next, we study the orthorhombic low-temperature structure of the system under external pressure. In Fig. \ref{distortionvsP} we show the evolution of the orthorhombic distortion at 15 K obtained from our x-ray synchrotron measurements on p1. We find that this distortion is suppressed by the application of external pressure and eventually disappears at $P_{c1} \simeq 9.4 $ GPa. This is similar to the observed in most of Fe-based SCs. Interestingly, by further increasing the pressure the structural distortion re-emerges at $P_{c2}\simeq 12.7$ GPa. This behavior strongly resembles the observed in LaFeAsO by means of H doping \cite{hosono-nphys14}.

\begin{table}[b!] 
\footnotesize
\begin{tabular}{ccccc}
\hline \hline
 & checkerboard & FM  & double-stripe & single-stripe \\
\hline 
$\Delta E$ (meV/Fe) & -5.71 & -11.11 & -11.26 & {\bf -44.56} \\
$\mu_{\rm Fe}$ ($\mu_{\rm B}$) & 0.90 & 0.65  & 1.04 & 1.16 \\
\hline \hline 
\end{tabular} 
\caption{Energy difference with respect to the non-magnetic state and Fe magnetic moments for various magnetic configurations obtained from DFT calculations. 
\label{table.energiesP=0}
}
\end{table}

\begin{figure}[b!]
\centering
\includegraphics[width=0.4\textwidth]{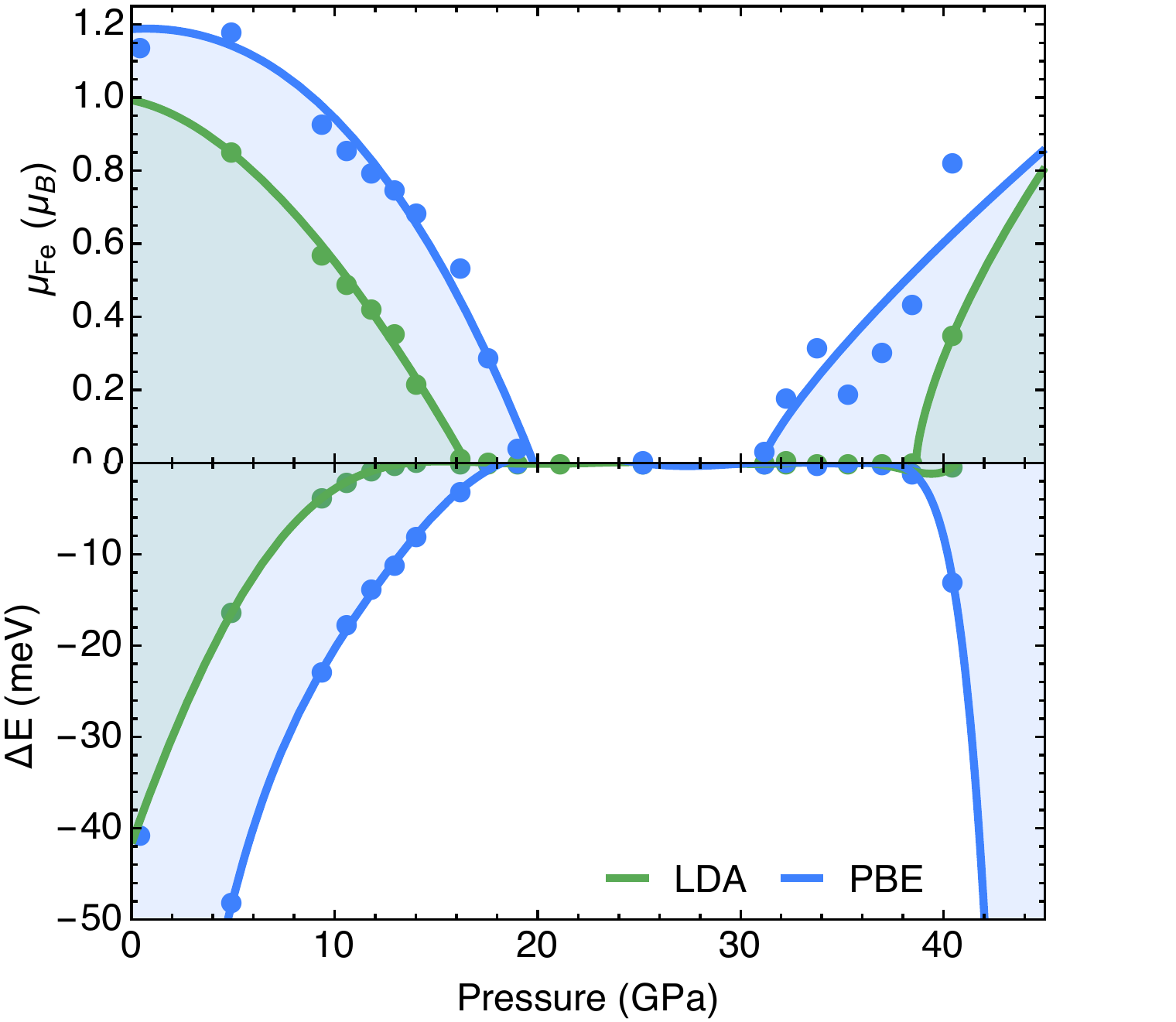}
	\caption{\label{AFMvsP} 
	Calculated Fe magnetic moment in the single-stripe AFM state (top) and energy difference with respect to the non-magnetic state (bottom) as a function of pressure. The results obtained with LDA and PBE functionals are represented by blue and green points respectively. Lines are guides to the eye.}
\end{figure}

In order to elucidate the nature of the high-pressure state we performed additional DFT calculations as a function of pressure for the aforementioned magnetic configurations. In these calculations, the single-stripe AFM state was always found to be the lowest-energy state. In Fig. \ref{AFMvsP} we show the resulting Fe magnetic moment and the energy difference with respect to the non-magnetic state as a function of pressure. Although the precise numerical values depend on the DFT method, they are in remarkably good agreement with the experimental observations. Namely, the application of external pressure first produces the suppression of single-stripe AFM order which then re-emerges as the pressure is further increased. Moreover, according to these results, the re-emergence is expected to be discontinuous. The reason is that, even if a non-vanishing Fe magnetic moment is developed, the energy of the single-stripe AFM state requires some extra pressure to eventually become lower than the energy of the non-magnetic state. The structural distortion observed in our experiments, however, displays rather continuous behavior. Note that this distortion is just a by-product of the magnetic order induced by magnetostructual coupling, and hence is quite small. Consequently, the magnetic discontinuity expected from the DFT calculations could be very difficult to observe indirectly via the lattice distortion. In any case, these calculations are remarkably consistent with the structural behavior observed in our x-ray experiments (see Fig. \ref{distortionvsP}).


\clearpage

\begin{figure}[t!]
\small \hspace{-0.4\textwidth}(a)\\
\includegraphics[height=0.34\textwidth]{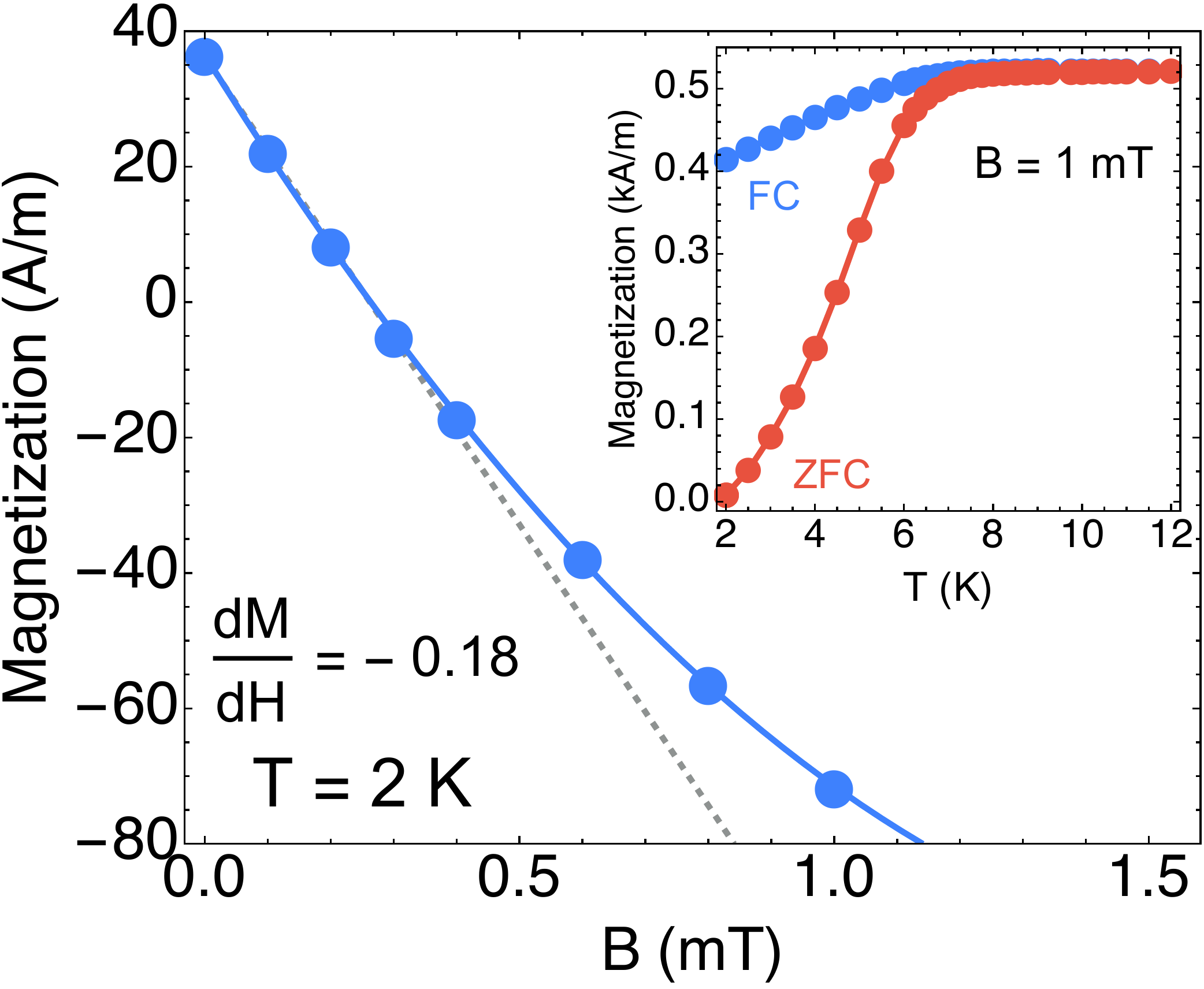}
\\
\small \hspace{-0.4\textwidth}(b)\\
\includegraphics[height=0.33\textwidth]{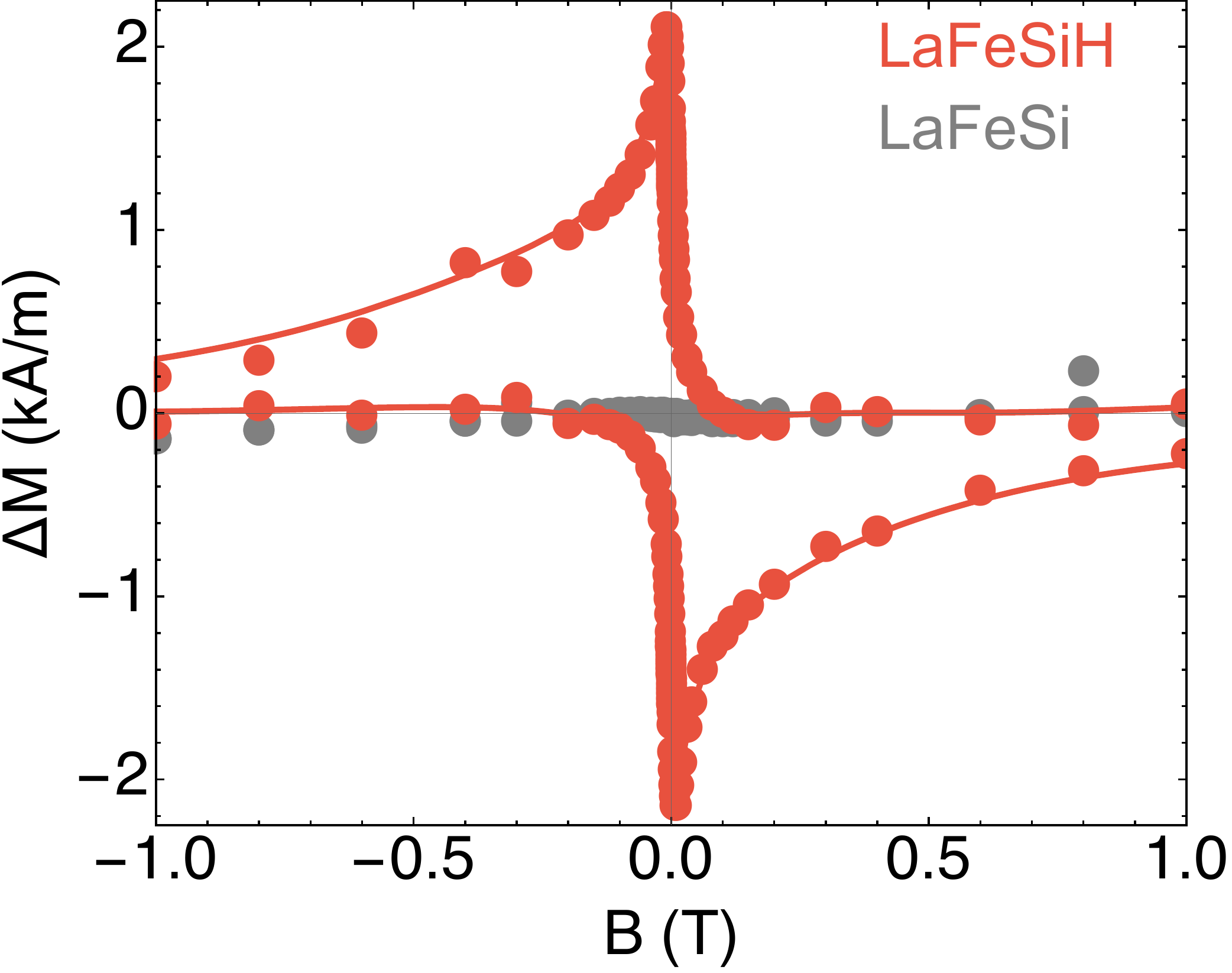}
\caption{
\blue{(a) Magnetization as a function of the external field at 2 K in p1. The diamagnetic response $dM/dH = -0.18$ evidences superconductivity in LaFeSiH. Inset: magnetization as a function of temperature at $B$ = 1 mT in p2, revealing a $\sim 64$~\% superconductor volume fraction in this powder sample. 
(b)} Superconducting hysteresis loop obtained from the magnetization loop at $2$\;K\;(below $T_c$) by substracting by the loop at 10\;K\;(above $>T_c$) (p2). The gray points correspond to the same subtraction for the non-hydrogenated original LaFeSi powder, which contains the same secondary ferromagnetic phases but is not superconductor (and hence displays no hysteresis).
\label{RandM}}
\end{figure}

\blue{
\paragraph{Electrical resistivity and magnetization.---}

Fig. \ref{rhovsH} displays the electrical resistivity measured in LaFeSiH single crystal (s2), which reveals the emergence of superconductivity in this novel compound. Specifically, the resistivity shows a $T^2$ Fermi-liquid behavior before the onset of superconductivity at 11~K [see also Fig. S6 in Supplemental Material]. From the drop in the resistivity to 50\% of its value at the onset we find the superconducting transition temperature $T_c = 9.7$~K at zero field. The inset in Fig. \ref{rhovsH} displays the values of the upper critical field $H_{c2}$ determined from the dependence of $T_c$ on the magnetic field applied in the $ab$ plane that can be seen in the main figure.
Using these values and the Werthamer-Helfand-Hohenberg formula $H_{c2}(0) = -0.69 \,T_c \left. d H_{c2}(T)/dT \right|_{T_c}$ we find $H_{c2}(0) \simeq 17$~T and the zero-temperature correlation length \blue{$\xi(0) \equiv \{\Phi_0/[2 \pi H_{c2}(0)]\}^{1/2} \simeq 4.3$~nm}. Note that the LaFeSiH $T_c$ cannot be explained in terms of conventional electron-phonon mediated superconductivity since it requires an unphysical $\mu^* =0$ Coulomb pseudopotential \cite{yildrim17}. 

In addition, we used the the powder sample p1 to determine the onset of superconductivity as a function of pressure from similar measurements [see Fig. S5(b)]. We find that pressure suppresses this onset in a rather smooth fashion as can be seen in Fig. \ref{TcvsP}. This behavior has a resemblance to the observed for the orthorhombic distortion [see Fig. \ref{distortionvsP}], and hence suggests a strong interplay between the corresponding instabilities. The re-emergence observed in the orthorhombic distortion, however, is not observed for the superconductivity below 21 GPa.

Fig. \ref{RandM} (a) shows the magnetization as a function of the external field measured in the powder p1. The initial magnetization is due the presence of the secondary ferromagnetic phases $\alpha$-Fe and La(Fe,Si)$_{13}$H$_x$ and depends on the history of the sample.
However, we note that the slope of the curve is negative and can be as strong as $dM/dH \sim -0.18$. This implies a global diamagnetic response whose strength is $4.4\times 10^2$ higher than that of pyrolytic carbon. This gives an lower-bound value for the superconducting diamagnetic strength of LaFeSiH, since it contains the sizable contribution of the aforementioned ferromagnetic phases. For the same reason, the value $H_{c1} \sim 0.3 $ mT that can be deduced for the lower critical field from the change in the slope has to be taken as a lower-bound value.
The inset in Fig. \ref{RandM} (a) shows the magnetization as a function of temperature measured in the powder p2 (see Supplemental Material for additional data in p1). The change observed in the ZFC curve between 10\;K and 2\;K reveals a superconductor volume fraction of $\sim 64$~\% in this powder. 

\begin{figure}[t!]
\includegraphics[height=0.33\textwidth]{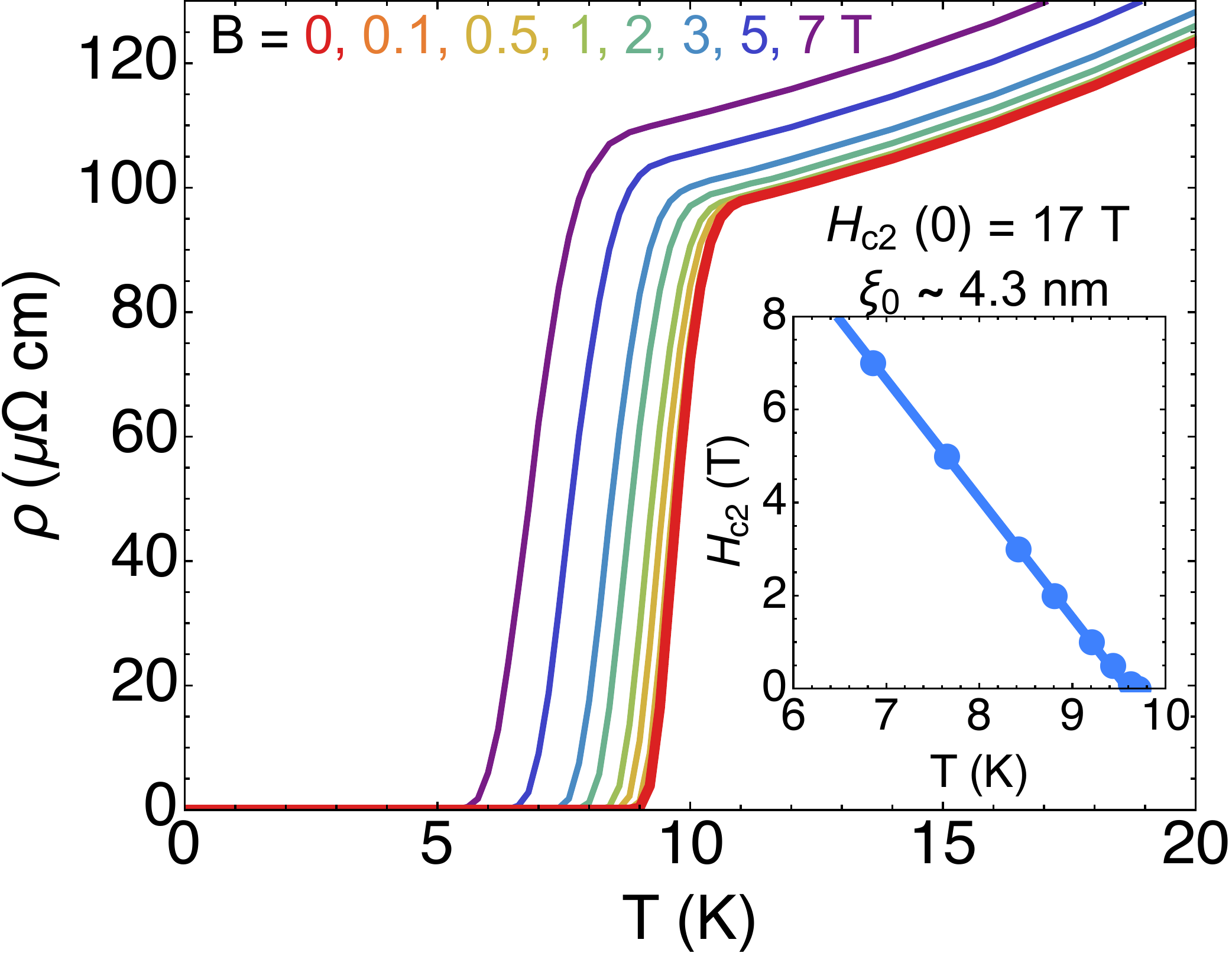}
\caption{
\blue{Low-temperature resisitivity as a function of temperature measured in single crystal LaFeSiH (s2) for different values of the magnetic field applied in the $ab$ plane. The inset shows the upper critical field as a function of temperature obtained from this data.} 
\label{rhovsH}}
\end{figure}
\begin{figure}[t!]
\hspace{5pt}\includegraphics[height=0.24\textwidth]{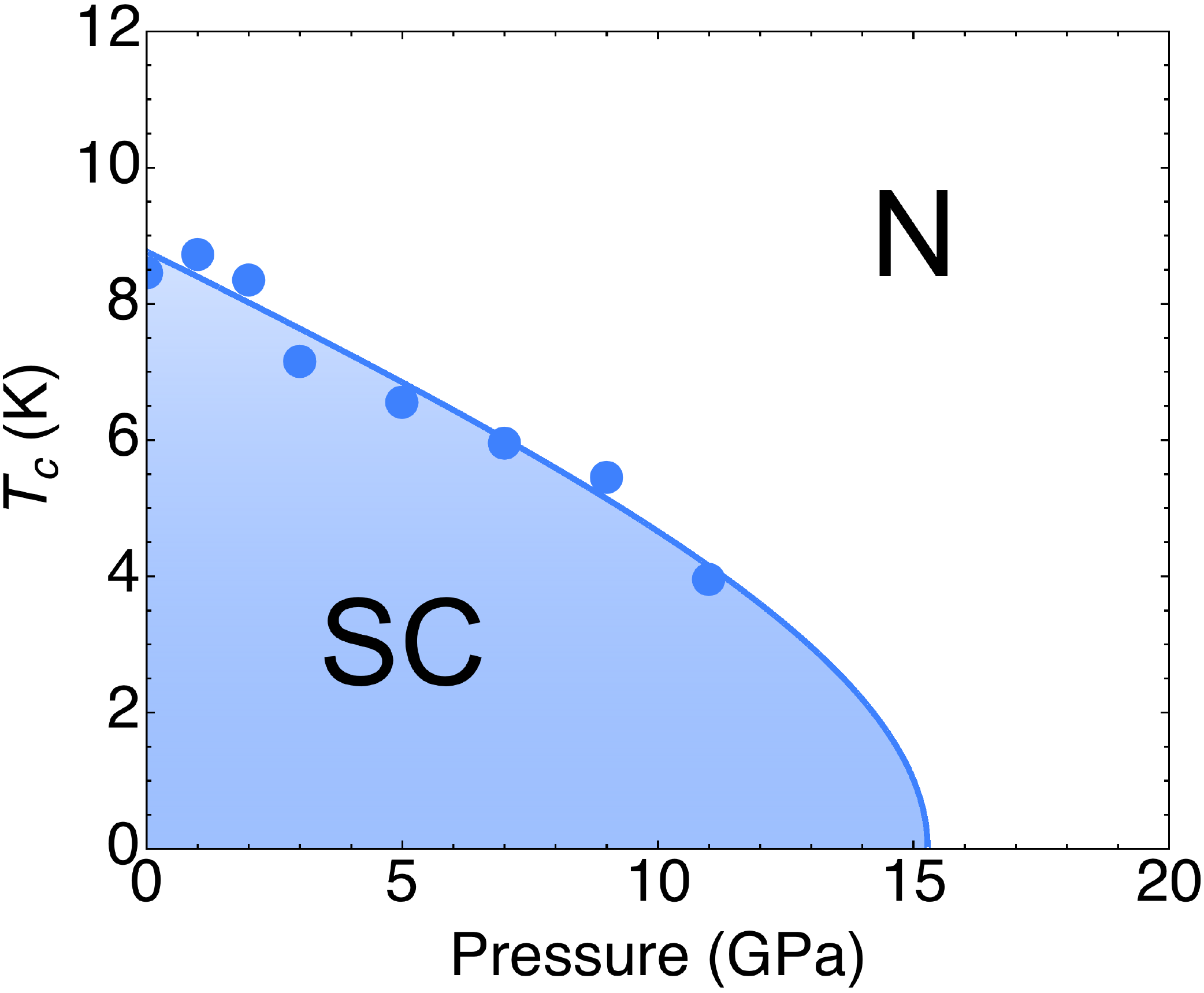}
\hspace{5pt}
\caption{
\blue{Superconducting onset as a function of pressure obtained from the resistance of the powder p1.}
\label{TcvsP}}
\end{figure}

We note that the ferromagnetic signal in our powders sets in at a much higher temperature \{$T_\text{FM}=1044$\,K for $\alpha$-Fe \cite{alphaFe} and 233-336\;K for La(Fe,Si)$_{13}$H$_x$ \cite{fujita03}\} and therefore is essentially temperature independent in the low-temperature regime (see Fig. S4 in Supplemental Material). Thus, by subtracting the magnetic loop obtained at 10\,K from the one at 2\,K we can obtain the remaining contribution due to the superconducting LaFeSiH. The resulting hysteresis loop is shown in Fig. \ref{RandM}(b). 

}

\paragraph{Conclusions.---} 
\blue{We have introduced FeSi as a new building block for iron-based superconductivity, which inaugurates a supplementary track to understanding the rich physics of high-temperature superconductors. We have reported the synthesis of the novel \blue{silicide hydride} LaFeSiH displaying superconductivity with onset 11 K. In addition, this silicide hydride} 
displays structural, magnetic, and electronic features \blue{similar to} previously reported \blue{iron-based superconductors}. Namely, a low-temperature tetragonal to orthorhombic distortion likely related to antiferromagnetic order and a multiband quasi-2D Fermi surface largely dominated by Fe-$3d$ orbitals. LaFeSiH \blue{therefore} demonstrates the possibility of iron-based superconductivity in a pnictogen- and chalcogen-free fashion. Thus, beyond its fundamental significance from both chemistry and physics point of view, LaFeSiH is expected to stimulate further research towards practical applications of iron-based superconducting materials.  

\blue{We thank ESRF and ILL for provision of beamtime at ID27 and D1B respectively. We also thank V. Nassif and O. Isnard for help with the neutron experiment, and M. Mezouar for fruitful discussions.}
F. B. acknowledges partial support from the FP7 European project SUPER-IRON (grant agreement No. 283204), and Progetto 
biennale d'ateneo UniCA/FdS/RAS CUP F72F16003050002. A.C. acknowledges support from French Government ``Investments for the Future'' Program, University of Bordeaux Initiative of Excellence (IDEX Bordeaux), and Sardinia Regional Government ``Visiting Professor'' Program 2015, University of Cagliari.


\begin{thebibliography}{30}%
\makeatletter
\providecommand \@ifxundefined [1]{%
 \@ifx{#1\undefined}
}%
\providecommand \@ifnum [1]{%
 \ifnum #1\expandafter \@firstoftwo
 \else \expandafter \@secondoftwo
 \fi
}%
\providecommand \@ifx [1]{%
 \ifx #1\expandafter \@firstoftwo
 \else \expandafter \@secondoftwo
 \fi
}%
\providecommand \natexlab [1]{#1}%
\providecommand \enquote  [1]{``#1''}%
\providecommand \bibnamefont  [1]{#1}%
\providecommand \bibfnamefont [1]{#1}%
\providecommand \citenamefont [1]{#1}%
\providecommand \href@noop [0]{\@secondoftwo}%
\providecommand \href [0]{\begingroup \@sanitize@url \@href}%
\providecommand \@href[1]{\@@startlink{#1}\@@href}%
\providecommand \@@href[1]{\endgroup#1\@@endlink}%
\providecommand \@sanitize@url [0]{\catcode `\\12\catcode `\$12\catcode
  `\&12\catcode `\#12\catcode `\^12\catcode `\_12\catcode `\%12\relax}%
\providecommand \@@startlink[1]{}%
\providecommand \@@endlink[0]{}%
\providecommand \url  [0]{\begingroup\@sanitize@url \@url }%
\providecommand \@url [1]{\endgroup\@href {#1}{\urlprefix }}%
\providecommand \urlprefix  [0]{URL }%
\providecommand \Eprint [0]{\href }%
\providecommand \doibase [0]{http://dx.doi.org/}%
\providecommand \selectlanguage [0]{\@gobble}%
\providecommand \bibinfo  [0]{\@secondoftwo}%
\providecommand \bibfield  [0]{\@secondoftwo}%
\providecommand \translation [1]{[#1]}%
\providecommand \BibitemOpen [0]{}%
\providecommand \bibitemStop [0]{}%
\providecommand \bibitemNoStop [0]{.\EOS\space}%
\providecommand \EOS [0]{\spacefactor3000\relax}%
\providecommand \BibitemShut  [1]{\csname bibitem#1\endcsname}%
\let\auto@bib@innerbib\@empty
\bibitem [{\citenamefont {{Martinelli}}\ \emph {et~al.}(2016)\citenamefont
  {{Martinelli}}, \citenamefont {{Bernardini}},\ and\ \citenamefont
  {{Massidda}}}]{martinelli-crp}%
  \BibitemOpen
  \bibfield  {author} {\bibinfo {author} {\bibfnamefont {A.}~\bibnamefont
  {{Martinelli}}}, \bibinfo {author} {\bibfnamefont {F.}~\bibnamefont
  {{Bernardini}}}, \ and\ \bibinfo {author} {\bibfnamefont {S.}~\bibnamefont
  {{Massidda}}},\ }\href {\doibase 10.1016/j.crhy.2015.06.001} {\bibfield
  {journal} {\bibinfo  {journal} {Comptes Rendus Physique}\ }\textbf {\bibinfo
  {volume} {17}},\ \bibinfo {pages} {5} (\bibinfo {year} {2016})}\BibitemShut
  {NoStop}%
\bibitem [{\citenamefont {{Bascones}}\ \emph {et~al.}(2016)\citenamefont
  {{Bascones}}, \citenamefont {{Valenzuela}},\ and\ \citenamefont
  {{Calder{\'o}n}}}]{bascones-crp}%
  \BibitemOpen
  \bibfield  {author} {\bibinfo {author} {\bibfnamefont {E.}~\bibnamefont
  {{Bascones}}}, \bibinfo {author} {\bibfnamefont {B.}~\bibnamefont
  {{Valenzuela}}}, \ and\ \bibinfo {author} {\bibfnamefont {M.~J.}\
  \bibnamefont {{Calder{\'o}n}}},\ }\href {\doibase 10.1016/j.crhy.2015.05.004}
  {\bibfield  {journal} {\bibinfo  {journal} {Comptes Rendus Physique}\
  }\textbf {\bibinfo {volume} {17}},\ \bibinfo {pages} {36} (\bibinfo {year}
  {2016})}\BibitemShut {NoStop}%
\bibitem [{\citenamefont {{Inosov}}(2016)}]{inosov-crp}%
  \BibitemOpen
  \bibfield  {author} {\bibinfo {author} {\bibfnamefont {D.~S.}\ \bibnamefont
  {{Inosov}}},\ }\href {\doibase 10.1016/j.crhy.2015.03.001} {\bibfield
  {journal} {\bibinfo  {journal} {Comptes Rendus Physique}\ }\textbf {\bibinfo
  {volume} {17}},\ \bibinfo {pages} {60} (\bibinfo {year} {2016})}\BibitemShut
  {NoStop}%
\bibitem [{\citenamefont {{B{\"o}hmer}}\ and\ \citenamefont
  {{Meingast}}(2016)}]{bohmer-crp}%
  \BibitemOpen
  \bibfield  {author} {\bibinfo {author} {\bibfnamefont {A.~E.}\ \bibnamefont
  {{B{\"o}hmer}}}\ and\ \bibinfo {author} {\bibfnamefont {C.}~\bibnamefont
  {{Meingast}}},\ }\href {\doibase 10.1016/j.crhy.2015.07.001} {\bibfield
  {journal} {\bibinfo  {journal} {Comptes Rendus Physique}\ }\textbf {\bibinfo
  {volume} {17}},\ \bibinfo {pages} {90} (\bibinfo {year} {2016})}\BibitemShut
  {NoStop}%
\bibitem [{\citenamefont {{van Roekeghem}}\ \emph {et~al.}(2016)\citenamefont
  {{van Roekeghem}}, \citenamefont {{Richard}}, \citenamefont {{Ding}},\ and\
  \citenamefont {{Biermann}}}]{roekeghem-crp}%
  \BibitemOpen
  \bibfield  {author} {\bibinfo {author} {\bibfnamefont {A.}~\bibnamefont {{van
  Roekeghem}}}, \bibinfo {author} {\bibfnamefont {P.}~\bibnamefont
  {{Richard}}}, \bibinfo {author} {\bibfnamefont {H.}~\bibnamefont {{Ding}}}, \
  and\ \bibinfo {author} {\bibfnamefont {S.}~\bibnamefont {{Biermann}}},\
  }\href {\doibase 10.1016/j.crhy.2015.11.003} {\bibfield  {journal} {\bibinfo
  {journal} {Comptes Rendus Physique}\ }\textbf {\bibinfo {volume} {17}},\
  \bibinfo {pages} {140} (\bibinfo {year} {2016})}\BibitemShut {NoStop}%
\bibitem [{\citenamefont {{Rullier-Albenque}}(2016)}]{rullier-crp}%
  \BibitemOpen
  \bibfield  {author} {\bibinfo {author} {\bibfnamefont {F.}~\bibnamefont
  {{Rullier-Albenque}}},\ }\href {\doibase 10.1016/j.crhy.2015.10.007}
  {\bibfield  {journal} {\bibinfo  {journal} {Comptes Rendus Physique}\
  }\textbf {\bibinfo {volume} {17}},\ \bibinfo {pages} {164} (\bibinfo {year}
  {2016})}\BibitemShut {NoStop}%
\bibitem [{\citenamefont {{Zapf}}\ \emph {et~al.}(2016)\citenamefont {{Zapf}},
  \citenamefont {{Neubauer}}, \citenamefont {{Post}}, \citenamefont {{Kadau}},
  \citenamefont {{Merz}}, \citenamefont {{Clauss}}, \citenamefont
  {{L{\"o}hle}}, \citenamefont {{Jeevan}}, \citenamefont {{Gegenwart}},
  \citenamefont {{Basov}},\ and\ \citenamefont {{Dressel}}}]{zapf-crp}%
  \BibitemOpen
  \bibfield  {author} {\bibinfo {author} {\bibfnamefont {S.}~\bibnamefont
  {{Zapf}}}, \bibinfo {author} {\bibfnamefont {D.}~\bibnamefont {{Neubauer}}},
  \bibinfo {author} {\bibfnamefont {K.~W.}\ \bibnamefont {{Post}}}, \bibinfo
  {author} {\bibfnamefont {A.}~\bibnamefont {{Kadau}}}, \bibinfo {author}
  {\bibfnamefont {J.}~\bibnamefont {{Merz}}}, \bibinfo {author} {\bibfnamefont
  {C.}~\bibnamefont {{Clauss}}}, \bibinfo {author} {\bibfnamefont
  {A.}~\bibnamefont {{L{\"o}hle}}}, \bibinfo {author} {\bibfnamefont {H.~S.}\
  \bibnamefont {{Jeevan}}}, \bibinfo {author} {\bibfnamefont {P.}~\bibnamefont
  {{Gegenwart}}}, \bibinfo {author} {\bibfnamefont {D.~N.}\ \bibnamefont
  {{Basov}}}, \ and\ \bibinfo {author} {\bibfnamefont {M.}~\bibnamefont
  {{Dressel}}},\ }\href {\doibase 10.1016/j.crhy.2015.04.008} {\bibfield
  {journal} {\bibinfo  {journal} {Comptes Rendus Physique}\ }\textbf {\bibinfo
  {volume} {17}},\ \bibinfo {pages} {188} (\bibinfo {year} {2016})}\BibitemShut
  {NoStop}%
\bibitem [{\citenamefont {{Hirschfeld}}(2016)}]{hirschfeld-crp}%
  \BibitemOpen
  \bibfield  {author} {\bibinfo {author} {\bibfnamefont {P.~J.}\ \bibnamefont
  {{Hirschfeld}}},\ }\href {\doibase 10.1016/j.crhy.2015.10.002} {\bibfield
  {journal} {\bibinfo  {journal} {Comptes Rendus Physique}\ }\textbf {\bibinfo
  {volume} {17}},\ \bibinfo {pages} {197} (\bibinfo {year} {2016})}\BibitemShut
  {NoStop}%
\bibitem [{\citenamefont {{Gallais}}\ and\ \citenamefont
  {{Paul}}(2016)}]{gallais-crp}%
  \BibitemOpen
  \bibfield  {author} {\bibinfo {author} {\bibfnamefont {Y.}~\bibnamefont
  {{Gallais}}}\ and\ \bibinfo {author} {\bibfnamefont {I.}~\bibnamefont
  {{Paul}}},\ }\href {\doibase 10.1016/j.crhy.2015.10.001} {\bibfield
  {journal} {\bibinfo  {journal} {Comptes Rendus Physique}\ }\textbf {\bibinfo
  {volume} {17}},\ \bibinfo {pages} {113} (\bibinfo {year} {2016})}\BibitemShut
  {NoStop}%
\bibitem [{\citenamefont {Yildirim}(2008)}]{yildrim-prl08}%
  \BibitemOpen
  \bibfield  {author} {\bibinfo {author} {\bibfnamefont {T.}~\bibnamefont
  {Yildirim}},\ }\href {\doibase 10.1103/PhysRevLett.101.057010} {\bibfield
  {journal} {\bibinfo  {journal} {Phys. Rev. Lett.}\ }\textbf {\bibinfo
  {volume} {101}},\ \bibinfo {pages} {057010} (\bibinfo {year}
  {2008})}\BibitemShut {NoStop}%
\bibitem [{\citenamefont {{Cano}}\ \emph {et~al.}(2010)\citenamefont {{Cano}},
  \citenamefont {{Civelli}}, \citenamefont {{Eremin}},\ and\ \citenamefont
  {{Paul}}}]{cano-prb10}%
  \BibitemOpen
  \bibfield  {author} {\bibinfo {author} {\bibfnamefont {A.}~\bibnamefont
  {{Cano}}}, \bibinfo {author} {\bibfnamefont {M.}~\bibnamefont {{Civelli}}},
  \bibinfo {author} {\bibfnamefont {I.}~\bibnamefont {{Eremin}}}, \ and\
  \bibinfo {author} {\bibfnamefont {I.}~\bibnamefont {{Paul}}},\ }\href
  {\doibase 10.1103/PhysRevB.82.020408} {\bibfield  {journal} {\bibinfo
  {journal} {Phys. Rev. B}\ }\textbf {\bibinfo {volume} {82}},\ \bibinfo {eid}
  {020408} (\bibinfo {year} {2010})}\BibitemShut {NoStop}%
\bibitem [{\citenamefont {Paul}\ \emph {et~al.}(2011)\citenamefont {Paul},
  \citenamefont {Cano},\ and\ \citenamefont {Sengupta}}]{cano11}%
  \BibitemOpen
  \bibfield  {author} {\bibinfo {author} {\bibfnamefont {I.}~\bibnamefont
  {Paul}}, \bibinfo {author} {\bibfnamefont {A.}~\bibnamefont {Cano}}, \ and\
  \bibinfo {author} {\bibfnamefont {K.}~\bibnamefont {Sengupta}},\ }\href
  {\doibase 10.1103/PhysRevB.83.115109} {\bibfield  {journal} {\bibinfo
  {journal} {Phys. Rev. B}\ }\textbf {\bibinfo {volume} {83}},\ \bibinfo
  {pages} {115109} (\bibinfo {year} {2011})}\BibitemShut {NoStop}%
\bibitem [{\citenamefont {{Garbarino}}\ \emph {et~al.}(2008)\citenamefont
  {{Garbarino}}, \citenamefont {{Toulemonde}}, \citenamefont
  {{{\'A}lvarez-Murga}}, \citenamefont {{Sow}}, \citenamefont {{Mezouar}},\
  and\ \citenamefont {{N{\'u}{\~n}ez-Regueiro}}}]{gg-prb-2008}%
  \BibitemOpen
  \bibfield  {author} {\bibinfo {author} {\bibfnamefont {G.}~\bibnamefont
  {{Garbarino}}}, \bibinfo {author} {\bibfnamefont {P.}~\bibnamefont
  {{Toulemonde}}}, \bibinfo {author} {\bibfnamefont {M.}~\bibnamefont
  {{{\'A}lvarez-Murga}}}, \bibinfo {author} {\bibfnamefont {A.}~\bibnamefont
  {{Sow}}}, \bibinfo {author} {\bibfnamefont {M.}~\bibnamefont {{Mezouar}}}, \
  and\ \bibinfo {author} {\bibfnamefont {M.}~\bibnamefont
  {{N{\'u}{\~n}ez-Regueiro}}},\ }\href {\doibase 10.1103/PhysRevB.78.100507}
  {\bibfield  {journal} {\bibinfo  {journal} {Phys. Rev. B}\ }\textbf {\bibinfo
  {volume} {78}},\ \bibinfo {eid} {100507} (\bibinfo {year}
  {2008})}\BibitemShut {NoStop}%
\bibitem [{\citenamefont {{Garbarino}}\ \emph {et~al.}(2011)\citenamefont
  {{Garbarino}}, \citenamefont {{Weht}}, \citenamefont {{Sow}}, \citenamefont
  {{Sulpice}}, \citenamefont {{Toulemonde}}, \citenamefont
  {{{\'A}lvarez-Murga}}, \citenamefont {{Strobel}}, \citenamefont {{Bouvier}},
  \citenamefont {{Mezouar}},\ and\ \citenamefont
  {{N{\'u}{\~n}ez-Regueiro}}}]{gg-prb-2011}%
  \BibitemOpen
  \bibfield  {author} {\bibinfo {author} {\bibfnamefont {G.}~\bibnamefont
  {{Garbarino}}}, \bibinfo {author} {\bibfnamefont {R.}~\bibnamefont {{Weht}}},
  \bibinfo {author} {\bibfnamefont {A.}~\bibnamefont {{Sow}}}, \bibinfo
  {author} {\bibfnamefont {A.}~\bibnamefont {{Sulpice}}}, \bibinfo {author}
  {\bibfnamefont {P.}~\bibnamefont {{Toulemonde}}}, \bibinfo {author}
  {\bibfnamefont {M.}~\bibnamefont {{{\'A}lvarez-Murga}}}, \bibinfo {author}
  {\bibfnamefont {P.}~\bibnamefont {{Strobel}}}, \bibinfo {author}
  {\bibfnamefont {P.}~\bibnamefont {{Bouvier}}}, \bibinfo {author}
  {\bibfnamefont {M.}~\bibnamefont {{Mezouar}}}, \ and\ \bibinfo {author}
  {\bibfnamefont {M.}~\bibnamefont {{N{\'u}{\~n}ez-Regueiro}}},\ }\href
  {\doibase 10.1103/PhysRevB.84.024510} {\bibfield  {journal} {\bibinfo
  {journal} {Phys. Rev. B}\ }\textbf {\bibinfo {volume} {84}},\ \bibinfo {eid}
  {024510} (\bibinfo {year} {2011})}\BibitemShut {NoStop}%
\bibitem [{\citenamefont {{Welter}}\ \emph {et~al.}(1998)\citenamefont
  {{Welter}}, \citenamefont {{Malaman}},\ and\ \citenamefont
  {{Venturini}}}]{welter-98}%
  \BibitemOpen
  \bibfield  {author} {\bibinfo {author} {\bibfnamefont {R.}~\bibnamefont
  {{Welter}}}, \bibinfo {author} {\bibfnamefont {B.}~\bibnamefont {{Malaman}}},
  \ and\ \bibinfo {author} {\bibfnamefont {G.}~\bibnamefont {{Venturini}}},\
  }\href {\doibase 10.1016/S0038-1098(98)00473-6} {\bibfield  {journal}
  {\bibinfo  {journal} {Solid State Communications}\ }\textbf {\bibinfo
  {volume} {108}},\ \bibinfo {pages} {933} (\bibinfo {year}
  {1998})}\BibitemShut {NoStop}%
\bibitem [{\citenamefont {Liu}\ \emph {et~al.}(2012)\citenamefont {Liu},
  \citenamefont {Matsuishi}, \citenamefont {Fujitsu},\ and\ \citenamefont
  {Hosono}}]{hosono-MgFeGe-prb-12}%
  \BibitemOpen
  \bibfield  {author} {\bibinfo {author} {\bibfnamefont {X.}~\bibnamefont
  {Liu}}, \bibinfo {author} {\bibfnamefont {S.}~\bibnamefont {Matsuishi}},
  \bibinfo {author} {\bibfnamefont {S.}~\bibnamefont {Fujitsu}}, \ and\
  \bibinfo {author} {\bibfnamefont {H.}~\bibnamefont {Hosono}},\ }\href
  {\doibase 10.1103/PhysRevB.85.104403} {\bibfield  {journal} {\bibinfo
  {journal} {Phys. Rev. B}\ }\textbf {\bibinfo {volume} {85}},\ \bibinfo
  {pages} {104403} (\bibinfo {year} {2012})}\BibitemShut {NoStop}%
\bibitem [{\citenamefont {Jeschke}\ \emph {et~al.}(2013)\citenamefont
  {Jeschke}, \citenamefont {Mazin},\ and\ \citenamefont
  {Valent\'{\i}}}]{roser-MgFeGe-prb13}%
  \BibitemOpen
  \bibfield  {author} {\bibinfo {author} {\bibfnamefont {H.~O.}\ \bibnamefont
  {Jeschke}}, \bibinfo {author} {\bibfnamefont {I.~I.}\ \bibnamefont {Mazin}},
  \ and\ \bibinfo {author} {\bibfnamefont {R.}~\bibnamefont {Valent\'{\i}}},\
  }\href {\doibase 10.1103/PhysRevB.87.241105} {\bibfield  {journal} {\bibinfo
  {journal} {Phys. Rev. B}\ }\textbf {\bibinfo {volume} {87}},\ \bibinfo
  {pages} {241105} (\bibinfo {year} {2013})}\BibitemShut {NoStop}%
\bibitem [{\citenamefont {{Zou}}\ \emph {et~al.}(2014)\citenamefont {{Zou}},
  \citenamefont {{Feng}}, \citenamefont {{Logg}}, \citenamefont {{Chen}},
  \citenamefont {{Lampronti}},\ and\ \citenamefont
  {{Grosche}}}]{grosche-pss14}%
  \BibitemOpen
  \bibfield  {author} {\bibinfo {author} {\bibfnamefont {Y.}~\bibnamefont
  {{Zou}}}, \bibinfo {author} {\bibfnamefont {Z.}~\bibnamefont {{Feng}}},
  \bibinfo {author} {\bibfnamefont {P.~W.}\ \bibnamefont {{Logg}}}, \bibinfo
  {author} {\bibfnamefont {J.}~\bibnamefont {{Chen}}}, \bibinfo {author}
  {\bibfnamefont {G.}~\bibnamefont {{Lampronti}}}, \ and\ \bibinfo {author}
  {\bibfnamefont {F.~M.}\ \bibnamefont {{Grosche}}},\ }\href {\doibase
  10.1002/pssr.201409418} {\bibfield  {journal} {\bibinfo  {journal} {Physica
  Status Solidi Rapid Research Letters}\ }\textbf {\bibinfo {volume} {8}},\
  \bibinfo {pages} {928} (\bibinfo {year} {2014})}\BibitemShut {NoStop}%
\bibitem [{\citenamefont {Chen}\ \emph {et~al.}(2016)\citenamefont {Chen},
  \citenamefont {Semeniuk}, \citenamefont {Feng}, \citenamefont {Reiss},
  \citenamefont {Brown}, \citenamefont {Zou}, \citenamefont {Logg},
  \citenamefont {Lampronti},\ and\ \citenamefont {Grosche}}]{grosche-prl16}%
  \BibitemOpen
  \bibfield  {author} {\bibinfo {author} {\bibfnamefont {J.}~\bibnamefont
  {Chen}}, \bibinfo {author} {\bibfnamefont {K.}~\bibnamefont {Semeniuk}},
  \bibinfo {author} {\bibfnamefont {Z.}~\bibnamefont {Feng}}, \bibinfo {author}
  {\bibfnamefont {P.}~\bibnamefont {Reiss}}, \bibinfo {author} {\bibfnamefont
  {P.}~\bibnamefont {Brown}}, \bibinfo {author} {\bibfnamefont
  {Y.}~\bibnamefont {Zou}}, \bibinfo {author} {\bibfnamefont {P.~W.}\
  \bibnamefont {Logg}}, \bibinfo {author} {\bibfnamefont {G.~I.}\ \bibnamefont
  {Lampronti}}, \ and\ \bibinfo {author} {\bibfnamefont {F.~M.}\ \bibnamefont
  {Grosche}},\ }\href {\doibase 10.1103/PhysRevLett.116.127001} {\bibfield
  {journal} {\bibinfo  {journal} {Phys. Rev. Lett.}\ }\textbf {\bibinfo
  {volume} {116}},\ \bibinfo {pages} {127001} (\bibinfo {year}
  {2016})}\BibitemShut {NoStop}%
\bibitem [{\citenamefont {{Kim}}\ \emph {et~al.}(2015)\citenamefont {{Kim}},
  \citenamefont {{Ran}}, \citenamefont {{Mun}}, \citenamefont {{Hodovanets}},
  \citenamefont {{Tanatar}}, \citenamefont {{Prozorov}}, \citenamefont
  {{Bud'ko}},\ and\ \citenamefont {{Canfield}}}]{canfield-15}%
  \BibitemOpen
  \bibfield  {author} {\bibinfo {author} {\bibfnamefont {H.}~\bibnamefont
  {{Kim}}}, \bibinfo {author} {\bibfnamefont {S.}~\bibnamefont {{Ran}}},
  \bibinfo {author} {\bibfnamefont {E.~D.}\ \bibnamefont {{Mun}}}, \bibinfo
  {author} {\bibfnamefont {H.}~\bibnamefont {{Hodovanets}}}, \bibinfo {author}
  {\bibfnamefont {M.~A.}\ \bibnamefont {{Tanatar}}}, \bibinfo {author}
  {\bibfnamefont {R.}~\bibnamefont {{Prozorov}}}, \bibinfo {author}
  {\bibfnamefont {S.~L.}\ \bibnamefont {{Bud'ko}}}, \ and\ \bibinfo {author}
  {\bibfnamefont {P.~C.}\ \bibnamefont {{Canfield}}},\ }\href {\doibase
  10.1080/14786435.2015.1004378} {\bibfield  {journal} {\bibinfo  {journal}
  {Philosophical Magazine}\ }\textbf {\bibinfo {volume} {95}},\ \bibinfo
  {pages} {804} (\bibinfo {year} {2015})}\BibitemShut {NoStop}%
\bibitem [{\citenamefont {Iimura}\ \emph {et~al.}(2012)\citenamefont {Iimura},
  \citenamefont {Matsuishi}, \citenamefont {Sato}, \citenamefont {Hanna},
  \citenamefont {Muraba}, \citenamefont {Kim}, \citenamefont {Kim},
  \citenamefont {Takata},\ and\ \citenamefont {Hosono}}]{hosono-natcomm12}%
  \BibitemOpen
  \bibfield  {author} {\bibinfo {author} {\bibfnamefont {S.}~\bibnamefont
  {Iimura}}, \bibinfo {author} {\bibfnamefont {S.}~\bibnamefont {Matsuishi}},
  \bibinfo {author} {\bibfnamefont {H.}~\bibnamefont {Sato}}, \bibinfo {author}
  {\bibfnamefont {T.}~\bibnamefont {Hanna}}, \bibinfo {author} {\bibfnamefont
  {Y.}~\bibnamefont {Muraba}}, \bibinfo {author} {\bibfnamefont {S.~W.}\
  \bibnamefont {Kim}}, \bibinfo {author} {\bibfnamefont {J.~E.}\ \bibnamefont
  {Kim}}, \bibinfo {author} {\bibfnamefont {M.}~\bibnamefont {Takata}}, \ and\
  \bibinfo {author} {\bibfnamefont {H.}~\bibnamefont {Hosono}},\ }\href@noop {}
  {\bibfield  {journal} {\bibinfo  {journal} {Nat. Commun.}\ }\textbf {\bibinfo
  {volume} {3}},\ \bibinfo {pages} {943} (\bibinfo {year} {2012})}\BibitemShut
  {NoStop}%
\bibitem [{\citenamefont {{Hiraishi}}\ \emph {et~al.}(2014)\citenamefont
  {{Hiraishi}}, \citenamefont {{Iimura}}, \citenamefont {{Kojima}},
  \citenamefont {{Yamaura}}, \citenamefont {{Hiraka}}, \citenamefont {{Ikeda}},
  \citenamefont {{Miao}}, \citenamefont {{Ishikawa}}, \citenamefont {{Torii}},
  \citenamefont {{Miyazaki}}, \citenamefont {{Yamauchi}}, \citenamefont
  {{Koda}}, \citenamefont {{Ishii}}, \citenamefont {{Yoshida}}, \citenamefont
  {{Mizuki}}, \citenamefont {{Kadono}}, \citenamefont {{Kumai}}, \citenamefont
  {{Kamiyama}}, \citenamefont {{Otomo}}, \citenamefont {{Murakami}},
  \citenamefont {{Matsuishi}},\ and\ \citenamefont
  {{Hosono}}}]{hosono-nphys14}%
  \BibitemOpen
  \bibfield  {author} {\bibinfo {author} {\bibfnamefont {M.}~\bibnamefont
  {{Hiraishi}}}, \bibinfo {author} {\bibfnamefont {S.}~\bibnamefont
  {{Iimura}}}, \bibinfo {author} {\bibfnamefont {K.~M.}\ \bibnamefont
  {{Kojima}}}, \bibinfo {author} {\bibfnamefont {J.}~\bibnamefont {{Yamaura}}},
  \bibinfo {author} {\bibfnamefont {H.}~\bibnamefont {{Hiraka}}}, \bibinfo
  {author} {\bibfnamefont {K.}~\bibnamefont {{Ikeda}}}, \bibinfo {author}
  {\bibfnamefont {P.}~\bibnamefont {{Miao}}}, \bibinfo {author} {\bibfnamefont
  {Y.}~\bibnamefont {{Ishikawa}}}, \bibinfo {author} {\bibfnamefont
  {S.}~\bibnamefont {{Torii}}}, \bibinfo {author} {\bibfnamefont
  {M.}~\bibnamefont {{Miyazaki}}}, \bibinfo {author} {\bibfnamefont
  {I.}~\bibnamefont {{Yamauchi}}}, \bibinfo {author} {\bibfnamefont
  {A.}~\bibnamefont {{Koda}}}, \bibinfo {author} {\bibfnamefont
  {K.}~\bibnamefont {{Ishii}}}, \bibinfo {author} {\bibfnamefont
  {M.}~\bibnamefont {{Yoshida}}}, \bibinfo {author} {\bibfnamefont
  {J.}~\bibnamefont {{Mizuki}}}, \bibinfo {author} {\bibfnamefont
  {R.}~\bibnamefont {{Kadono}}}, \bibinfo {author} {\bibfnamefont
  {R.}~\bibnamefont {{Kumai}}}, \bibinfo {author} {\bibfnamefont
  {T.}~\bibnamefont {{Kamiyama}}}, \bibinfo {author} {\bibfnamefont
  {T.}~\bibnamefont {{Otomo}}}, \bibinfo {author} {\bibfnamefont
  {Y.}~\bibnamefont {{Murakami}}}, \bibinfo {author} {\bibfnamefont
  {S.}~\bibnamefont {{Matsuishi}}}, \ and\ \bibinfo {author} {\bibfnamefont
  {H.}~\bibnamefont {{Hosono}}},\ }\href {\doibase 10.1038/nphys2906}
  {\bibfield  {journal} {\bibinfo  {journal} {Nature Physics}\ }\textbf
  {\bibinfo {volume} {10}},\ \bibinfo {pages} {300} (\bibinfo {year}
  {2014})}\BibitemShut {NoStop}%
\bibitem [{\citenamefont {Welter}\ \emph {et~al.}(1998)\citenamefont {Welter},
  \citenamefont {Ijjaali}, \citenamefont {Venturini},\ and\ \citenamefont
  {Malaman}}]{welter-jac98}%
  \BibitemOpen
  \bibfield  {author} {\bibinfo {author} {\bibfnamefont {R.}~\bibnamefont
  {Welter}}, \bibinfo {author} {\bibfnamefont {I.}~\bibnamefont {Ijjaali}},
  \bibinfo {author} {\bibfnamefont {G.}~\bibnamefont {Venturini}}, \ and\
  \bibinfo {author} {\bibfnamefont {B.}~\bibnamefont {Malaman}},\ }\href
  {\doibase http://dx.doi.org/10.1016/S0925-8388(97)00280-6} {\bibfield
  {journal} {\bibinfo  {journal} {Journal of Alloys and Compounds}\ }\textbf
  {\bibinfo {volume} {265}},\ \bibinfo {pages} {196 } (\bibinfo {year}
  {1998})}\BibitemShut {NoStop}%
\bibitem [{\citenamefont {Tenc\'{e}}\ \emph {et~al.}(2009)\citenamefont
  {Tenc\'{e}}, \citenamefont {Andr\'{e}}, \citenamefont {Gaudin}, \citenamefont
  {Bonville}, \citenamefont {Al~Alam}, \citenamefont {Matar}, \citenamefont
  {Hermes}, \citenamefont {Pottgen},\ and\ \citenamefont
  {Chevalier}}]{tence-jap09}%
  \BibitemOpen
  \bibfield  {author} {\bibinfo {author} {\bibfnamefont {S.}~\bibnamefont
  {Tenc\'{e}}}, \bibinfo {author} {\bibfnamefont {G.}~\bibnamefont
  {Andr\'{e}}}, \bibinfo {author} {\bibfnamefont {E.}~\bibnamefont {Gaudin}},
  \bibinfo {author} {\bibfnamefont {P.}~\bibnamefont {Bonville}}, \bibinfo
  {author} {\bibfnamefont {A.~F.}\ \bibnamefont {Al~Alam}}, \bibinfo {author}
  {\bibfnamefont {S.~F.}\ \bibnamefont {Matar}}, \bibinfo {author}
  {\bibfnamefont {W.}~\bibnamefont {Hermes}}, \bibinfo {author} {\bibfnamefont
  {R.}~\bibnamefont {Pottgen}}, \ and\ \bibinfo {author} {\bibfnamefont
  {B.}~\bibnamefont {Chevalier}},\ }\href {\doibase
  http://dx.doi.org/10.1063/1.3190488} {\bibfield  {journal} {\bibinfo
  {journal} {Journal of Applied Physics}\ }\textbf {\bibinfo {volume} {106}},\
  \bibinfo {eid} {033910} (\bibinfo {year} {2009})}\BibitemShut {NoStop}%
\bibitem [{che(2009)}]{chevalier09}%
  \BibitemOpen
  \href {\doibase http://dx.doi.org/10.1016/j.jallcom.2008.09.183} {\bibfield
  {journal} {\bibinfo  {journal} {Journal of Alloys and Compounds}\ }\textbf
  {\bibinfo {volume} {480}},\ \bibinfo {pages} {43 } (\bibinfo {year}
  {2009})}\BibitemShut {NoStop}%
\bibitem [{ten(2010)}]{tence10}%
  \BibitemOpen
  \href {\doibase 10.1021/ic902079u} {\bibfield  {journal} {\bibinfo  {journal}
  {Inorganic Chemistry}\ }\textbf {\bibinfo {volume} {49}},\ \bibinfo {pages}
  {4836} (\bibinfo {year} {2010})}\BibitemShut {NoStop}%
\bibitem [{\citenamefont {{Hung}}\ and\ \citenamefont
  {{Yildirim}}(2017)}]{yildrim17}%
  \BibitemOpen
  \bibfield  {author} {\bibinfo {author} {\bibfnamefont {L.}~\bibnamefont
  {{Hung}}}\ and\ \bibinfo {author} {\bibfnamefont {T.}~\bibnamefont
  {{Yildirim}}},\ }\href@noop {} {\  (\bibinfo {year} {2017})},\ \Eprint
  {http://arxiv.org/abs/1711.01764} {arXiv:1711.01764} \BibitemShut {NoStop}%
\bibitem [{\citenamefont {Cano}\ and\ \citenamefont {Paul}(2012)}]{cano-prb12}%
  \BibitemOpen
  \bibfield  {author} {\bibinfo {author} {\bibfnamefont {A.}~\bibnamefont
  {Cano}}\ and\ \bibinfo {author} {\bibfnamefont {I.}~\bibnamefont {Paul}},\
  }\href {\doibase 10.1103/PhysRevB.85.155133} {\bibfield  {journal} {\bibinfo
  {journal} {Phys. Rev. B}\ }\textbf {\bibinfo {volume} {85}},\ \bibinfo
  {pages} {155133} (\bibinfo {year} {2012})}\BibitemShut {NoStop}%
\bibitem [{\citenamefont {{Coey}}(2009)}]{alphaFe}%
  \BibitemOpen
  \bibfield  {author} {\bibinfo {author} {\bibfnamefont {J.~M.~D.}\
  \bibnamefont {{Coey}}},\ }\href@noop {} {\emph {\bibinfo {title} {Magnetism
  and Magnetic materials}}}\ (\bibinfo  {publisher} {Cambridge University
  press, New York},\ \bibinfo {year} {2009})\BibitemShut {NoStop}%
\bibitem [{\citenamefont {{Fujita}}\ \emph {et~al.}(2003)\citenamefont
  {{Fujita}}, \citenamefont {{Fujieda}}, \citenamefont {{Hasegawa}},\ and\
  \citenamefont {{Fukamichi}}}]{fujita03}%
  \BibitemOpen
  \bibfield  {author} {\bibinfo {author} {\bibfnamefont {A.}~\bibnamefont
  {{Fujita}}}, \bibinfo {author} {\bibfnamefont {S.}~\bibnamefont {{Fujieda}}},
  \bibinfo {author} {\bibfnamefont {Y.}~\bibnamefont {{Hasegawa}}}, \ and\
  \bibinfo {author} {\bibfnamefont {K.}~\bibnamefont {{Fukamichi}}},\ }\href
  {\doibase 10.1103/PhysRevB.67.104416} {\bibfield  {journal} {\bibinfo
  {journal} {\prb}\ }\textbf {\bibinfo {volume} {67}},\ \bibinfo {eid} {104416}
  (\bibinfo {year} {2003})}\BibitemShut {NoStop}%
\end{thebibliography}
%

\end{document}